\documentclass[aps,prl,twocolumn,amsmath,amssymb,floatfix,superscriptaddress]{revtex4-1}
\usepackage{tabularx}
\usepackage{bm}
\usepackage{euscript}
\usepackage{epsfig,psfrag,subfigure}
\usepackage{graphicx}
\usepackage{color}
\usepackage{amsfonts}
\usepackage{exscale}
\usepackage{amsbsy}
\usepackage{multirow}
\usepackage{hyperref}

\usepackage[dvipsnames]{xcolor}
\def\be{\begin{equation}}       \def\ee{\end{equation}}
\def\bea{\begin{eqnarray}}      \def\eea{\end{eqnarray}}
\def\trace#1{{\rm Tr}\left[#1\right]}
\newcommand{\PreserveBackslash}[1]{\let\temp=\\#1\let\\=\temp}
\newcolumntype{C}[1]{>{\PreserveBackslash\centering}p{#1}}

\begin{document}

\title{Evidence of pair-density wave in doping Kitaev spin liquid on the honeycomb lattice}
\author{Cheng Peng}
\thanks{These authors contributed equally.}
\affiliation{Stanford Institute for Materials and Energy Sciences, SLAC National Accelerator Laboratory and Stanford University, Menlo Park, California 94025, USA}

\author{Yi-Fan Jiang}
\thanks{These authors contributed equally.}
\affiliation{Stanford Institute for Materials and Energy Sciences, SLAC National Accelerator Laboratory and Stanford University, Menlo Park, California 94025, USA}

\author{Thomas P. Devereaux}
\affiliation{Stanford Institute for Materials and Energy Sciences, SLAC National Accelerator Laboratory and Stanford University, Menlo Park, California 94025, USA}
\affiliation{{Department of Materials Science and Engineering, Stanford University, Stanford, CA 94305, USA}}


\author{Hong-Chen Jiang}
\email{hcjiang@stanford.edu}
\affiliation{Stanford Institute for Materials and Energy Sciences, SLAC National Accelerator Laboratory and Stanford University, Menlo Park, California 94025, USA}

\begin{abstract}
We study the effects of doping the Kitaev model on the honeycomb lattice where the spins interact via the bond-directional interaction $J_K$, which is known to have a quantum spin liquid as its exact ground state. The effect of hole doping is studied within the $t$-$J_K$ model on a three-leg cylinder using density-matrix renormalization group. Upon light doping, we find that the ground state of the system has quasi-long-range charge-density-wave correlations but short-range 
single-particle correlations. The dominant pairing channel is the even-parity superconducting pair-pair correlations with $d$-wave-like symmetry, which oscillate in sign as a function of separation with a period equal to that of the spin-density wave and two times the charge-density wave. Although these correlations fall rapidly (possibly exponentially) at long distances, this is never-the-less the first example where a pair-density wave is the strongest SC order on a bipartite lattice.  Our results may be relevant to ${\rm Na_2IrO_3}$ and $\alpha$-${\rm RuCl_3}$ upon doping.
\end{abstract}

\maketitle

The pair-density wave (PDW) is a 
superconducting (SC) state in which the order parameter varies periodically in space in such a way that its spatial average vanishes.\cite{Berg2009,Agterberg2020} The first example of PDW is the Fulde-Ferrell-Larkin-Ovchinnikov (FFLO) state\cite{FF1964,LO1965} when a Zeeman magnetic field, $H$, is applied to a s-wave superconductor so that the Fermi surface is spin-split. The SC order has a wave vector $Q \sim \mu_BH/E_F$ which is typically very small. The LO version of this state is accompanied by an induced magnetization density wave and a charge density wave (CDW) with ordering wavevectors $K=2Q$. Intense interest in a somewhat different sort of PDW state has emerged due to recent discoveries in underdoped cuprate superconductors, where a direct observation of PDW has been made experimentally via local Cooper pair tunneling and scanning tunneling microscopy in ${\rm Bi_2Sr_2Ca_2O_{8+x}}$\cite{Hamidian2016,Ruan2018,Edkins2019} as well as the dynamical inter-layer decoupling observed in $1/8$ hole doped ${\rm{La}_2BaCuO_4}$.\cite{Berg2007,Agterberg2008} While similar in having  oscillatory SC order and associated $K=2Q$ CDW order, this PDW is conjectured to be stable in zero magnetic field (zero net magnetization), have an ordering vector that is independent of $H$ (at least for small or vanishing $H$), and moreover  can either have no associated magnetic order, or possibly have spin density wave order (SDW) with the same ordering vector $Q$.

Although much is known about the properties of the PDW state\cite{Berg2007,Agterberg2008,Berg2009,Lee2014,Jian2020}, there are very few microscopic models which are shown to have PDW ground states. These include the one-dimensional (1D) Kondo-Heisenberg model with 1D electron gas coupled to a spin chain\cite{Berg2010} and the extended two-leg Hubbard-Heisenberg model\cite{Fradkin2012}. The evidence of PDW is also observed in the t-J model with four-spin ring exchange on a 4-leg triangular lattice\cite{Xu2019} and an extended Hubbard model with a staggered spin-dependent magnetic flux per plaquette on a 3-leg triangular lattice.\cite{Venderley2019} However, there is no evidence of PDW ordering found in more standard models 
even with second neighbor interactions.\cite{Dodaro2017,Jiang2018tJ,Jiang2019hub,Jiang2020hub}

Theoretically, it has been proposed that superconductivity can also emerge in doping quantum spin liquids (QSLs), which are exotic phases of matter that exhibit a variety of novel features associated with their topological character.\cite{Balents2010,Savary2016,Broholm2019} The QSLs can be viewed as insulating phases with preexisting electron pairs such that upon light doping they might automatically yield high temperature superconductivity.\cite{Anderson1987,Kivelson1987,Rokhsar1988,Laughlin1988,Wen1989,Lee2006,Lee2007,Fradkin2015} Indeed, recent numerical studies using density-matrix renormalization group (DMRG) have provided strong evidences that lightly doping the QSL and chiral spin liquid on the triangular lattice will naturally give rise to nematic $d$-wave\cite{Jiang2019tqsl} and $d\pm id$-wave superconductivity\cite{Jiang2020tcsl}, respectively. Although dominant SC correlations have been observed in both systems, there is still no evidence for PDW ordering.

 \begin{figure}[tb]
      \includegraphics[angle=0,width=0.8\linewidth]{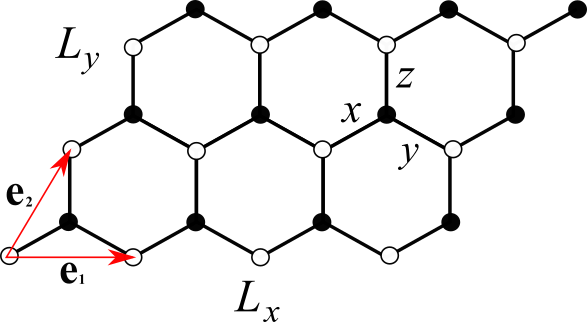}
      \caption{(Color online) The schematic three-leg cylinder on the honeycomb lattice. The open (filled) circle denotes $A$ ($B$) sub-lattice, and $x$, $y$ and $z$ label the three different bonds. Periodic (open) boundary condition is imposed along the direction specified by the lattice basis vector $\mathbf{e}_2$ ($\mathbf{e}_1$). $L_x$ ($L_y$) is the number of unit cells in the $\mathbf{e}_1$ ($\mathbf{e}_2$) direction.}\label{Fig:Lattice}
\end{figure}%

In this paper, we define a $t$-$J$-like extension of the Kitaev model on the honeycomb lattice (Fig.\ref{Fig:Lattice}) so as to address the question of whether SC emerges upon light doping. The Kitaev model, i.e., $J_K$ term in Eq.(\ref{Eq:Ham}), is exactly solvable and has a gapless spin liquid as its exact ground state.\cite{Kitaev2006} Moreover, it has potential experimental realizations in magnets with strong spin-orbit coupling such as ${\rm Na_2IrO_3}$ and ${\rm \alpha}$-${\rm RuCl_3}$.\cite{Jackeli2009,Jiang2011,Singh2012,Chun2015,Kim2016,Banerjee2017,Trebst2017,Takagi2019,Hickey2019,Jiang2019kh,Yokoi2020,Chern2020} This provides us a unique theoretical opportunity to test the physics of doping QSLs and may also give us some hints for understanding the mechanism of high temperature superconductiivity in the cuprates. Theoretically, doping Kitaev spin liquid (KSL) has been studied and distinct metallic states were proposed.\cite{You2012,Hyart2012,Schmidt2018,Mei2012} These include the $p$-wave superconductivity\cite{You2012,Hyart2012}, topological superconductivity\cite{Schmidt2018}, and Fermi liquid state.\cite{Mei2012} 
However, controlled results of the sort that can be obtained using density-matrix renormalization group (DMRG) are still lacking concerning the phase(s) that arise upon doping the KSL.

\textbf{Principal results:} %
In the present paper, we study the lightly doped Kitaev model in Eq.(\ref{Eq:Ham}) on the honeycomb lattice using DMRG\cite{White1992}. Based on DMRG calculations on three-leg cylinders, we find that upon light-doping the KSL state, the system exhibits power-law CDW correlations at long distances corresponding to a local pattern of partially-filled charge stripes. For three-leg cylinders, the wavelength of CDW, i.e., the spacings between two adjacent charge strips in the $\mathbf{e}_1$ direction, is $\lambda_{c}=a_0/3\delta$, where $a_0$ is the length of unit cell. This corresponds to an ordering wavevector $K=3\pi \delta/a_0$ with two thirds of a doped hole per CDW unit cell.

We find that the even-parity SC correlations are the most pronounced SC correlations, far dominant compared with the odd-parity SC or topological superconductivity.\cite{You2012,Hyart2012,Schmidt2018} Moreover, the dominant pairing channel oscillates in sign as a function of distance which is consistent with the striped PDW.\cite{Agterberg2020} Its wavelength $\lambda_{sc}=2a_0/3\delta$ is the same as that of the spin density wave (SDW) correlations $\lambda_s=2a_0/3\delta$ and two times of that of the CDW $\lambda_c=a_0/3\delta$. Correspondingly, the SC ordering wavevector $Q=3\pi\delta/2$ is half of the ordering wavevector $K=3\pi\delta/a_0$ of the CDW. Although quasi-long-range SC correlations are not seen in the range of parameters we have studied, short-range correlations are fairly strong with the corresponding correlation length $\xi_{sc}\geq 3a_0$. This is comparable with the cylindrical width and is notably larger than that of doping the Kagome QSL.\cite{Jiang2017kqsl} Similar with doping the QSL on the triangular lattice\cite{Jiang2019tqsl}, we find that the pairing symmetry of SC correlations is also consistent with $d$-wave. To the best of our knowledge, this is the first observation of PDW in doping the KSL.

\textbf{Model and Method: }%
We employ DMRG\cite{White1992} to investigate the ground state properties of the hole-doped Kitaev model on the honeycomb lattice defined by the Hamiltonian %
\begin{eqnarray}
  H=-t\sum_{\langle ij\rangle, \sigma}(c^{\dagger}_{i\sigma}c_{j\sigma}+h.c.) + J_K\sum_{\langle ij\rangle}S^{\gamma}_iS^{\gamma}_j.
  \label{Eq:Ham}
\end{eqnarray}
Here $c^{\dagger}_{i\sigma}$($c_{i\sigma}$) is the electron creation (annihilation) operator with spin-$\sigma$ on site $i=(x_i,y_i)$, $S^\gamma_i$ is the $\gamma$-component of the $S=1/2$ spin operator on site $i$, where $\gamma=x,y,z$ labels the three different links of the hexagonal lattice as illustrated in Fig.\ref{Fig:Lattice}. $\langle ij\rangle$ denotes nearest-neighbor (NN) sites and the Hilbert space is constrained by the no-double occupancy condition, $n_i\leq 1$, where $n_i=\sum_\sigma c^\dagger_{i\sigma}c_{i\sigma}$ is the electron number operator. At half-filling, i.e., $n_i=1$, Eq.(\ref{Eq:Ham}) reduces to the Kitaev model, which is known to have a gapless spin liquid ground state 
that can be gapped out into a non-Abelian topological phase by certain time-reversal symmetry perturbations.\cite{Kitaev2006,Zhu2018}

\begin{figure}[tb]
\centering
    \includegraphics[width=1\linewidth]{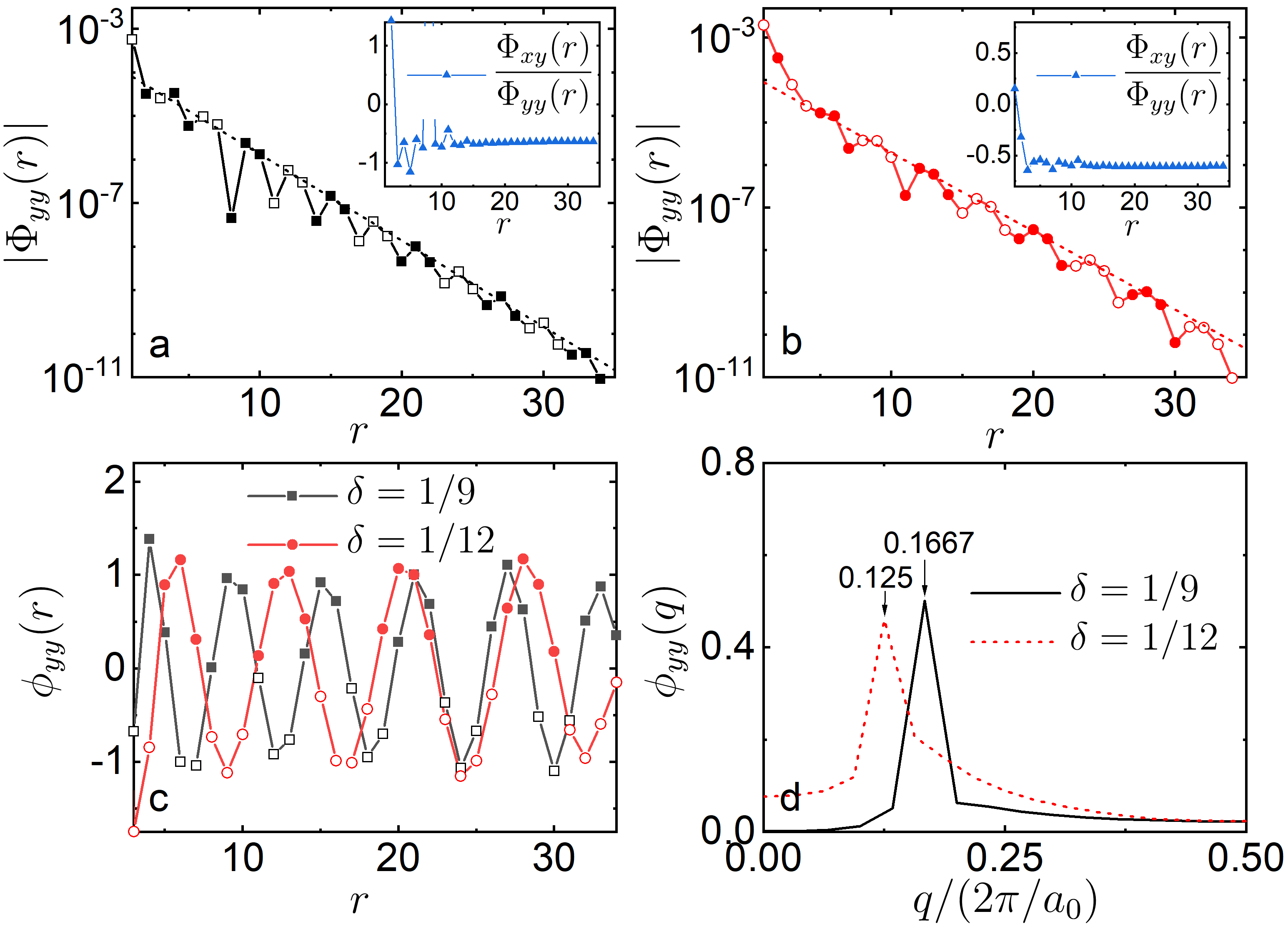}
\caption{(Color online) Superconducting correlations $|\Phi_{yy}(r)|$ on three-leg cylinder at doping level (a) $\delta=1/9$ and (b) $\delta=1/12$, where dashed lines denote fittings to an exponential function $f(r)\sim e^{-r/\xi_{sc}}$. Insets are the ratio $\Phi_{xy}(r)/\Phi_{yy}(r)<0$, where the out of phase indicates $d$-wave type pairing. (c) The normalized function $\phi_{yy}(r)=\Phi_{yy}(r)/f(r)$ at $\delta=1/9$ and $\delta=1/12$, which directly reflects the spatial oscillation of $\Phi_{yy}(r)$. (d) Fourier transformation $\phi_{yy}(q)$ of $\phi_{yy}(r)$ at $\delta=1/12$ and $\delta=1/9$, where peaks lie at $Q=3\pi\delta/2a_0$ giving the total momentum of the pairing. Note that filled (open) symbols denote positive (negative) value.} \label{Fig:PDW}
\end{figure}

The lattice geometry used in our simulations is depicted in Fig.\ref{Fig:Lattice}, where $\mathbf{e}_1=(\sqrt{3},0)$ and $\mathbf{e}_2=(1/2,\sqrt{3}/2)$ denote the two basis vectors. We consider honeycomb cylinders with periodic and open boundary conditions in the $\mathbf{e}_2$ and $\mathbf{e}_1$ directions, respectively. We focus on cylinders of width $L_y$ and length $L_x$, where $L_y$ and $L_x$ are the number of unit cells ($L_y$ and $\tilde{L}_x=2L_x$ are the number of sites) along the $\mathbf{e}_2$ and $\mathbf{e}_1$ directions, respectively. The total number of sites is $N=2\times L_y\times L_x$. The hole doping concentration is defined as $\delta=N_h/N$, where $N_h$ is the number of doped holes. We set $J_K=1$ as an energy unit 
and consider $t=3$. In this paper, we focus primarily on three-leg cylinders, i.e., $L_y=3a_0$, of length $L_x=12a_0\sim 48a_0$ (i.e., $\tilde{L}_x=24\sim 96$), at doping levels $\delta=1/12$ and $1/9$. We find similar results on four-leg cylinders 
which are provided in the Supplemental Materials (SM). We keep up to $m=8000$ number of states in each DMRG block with a typical truncation error $\epsilon\sim 10^{-6}$. This leads to excellent convergence for our results when extrapolated to $m=\infty$ limit.

\textbf{Pair-density wave: }%
To test the possibility of superconductivity, we have calculated the equal-time SC pair-pair correlations. A diagnostic of the SC order is the SC pair-pair correlation function, defined as
\begin{eqnarray}
\Phi_{\alpha\beta}(r)=\langle\Delta^{\dagger}_{\alpha}(x_0,y)\Delta_{\beta}(x_0+r,y)\rangle, 
\label{Eq:SC}
\end{eqnarray}
where $\Delta^{\dagger}_{\alpha}(x,y)=\frac{1}{\sqrt{2}}[\hat{c}^{\dagger}_{(x,y),\uparrow}\hat{c}^{\dagger}_{(x,y)+\alpha,\downarrow} - \hat{c}^{\dagger}_{(x,y),\downarrow}\hat{c}^{\dagger}_{(x,y)+\alpha,\uparrow}]$ is the even-parity SC pair-field creation operator, where the bond orientations are labelled as $\alpha=x,y,z$ (Fig.\ref{Fig:Lattice}). ($x_0,y$) is the reference bond with $x_0\sim \tilde{L}_x/4$ to minimize the boundary effect and $r$ is the distance between two bonds in the $\mathbf{e}_1$ direction. We have also calculated the odd-parity SC correlations to test the possibility of $p$-wave or topological superconductivity. However, we find that they are much weaker than the even-parity SC correlations (see SM for details), suggesting that $p$-wave or topological superconductivity is unlikely. Therefore, we will focus on the even-parity SC correlation in this paper.

Fig.\ref{Fig:PDW} shows the SC pair-pair function $\Phi_{yy}(r)$ for doping $\delta=1/12$ and $1/9$. The SC correlation shows clear spatial oscillation for both doping levels, which can be well fitted by $\Phi_{yy}(r)\sim f(r)*\phi_{yy}(r)$ for a large region of $r$ as we will discuss below. Here $f(r)$ is the envelope function and $\phi_{yy}(r)$ is a spatial oscillatory function. At long distances, the envelope function $f(r)$ is consistent with an exponential decay, i.e., $f(r)\sim e^{-r/\xi_{sc}}$, as shown in Fig.\ref{Fig:PDW}a-b. The extracted correlation length is $\xi_{sc}\geq 3a_0$, which is comparable with the cylindrical width $L_y=3a_0$. Alternatively, the SC correlation at long distances can also be fitted by a power law (see SM for details), i.e., $f(r)\sim r^{-K_{sc}}$, with an exponent $K_{sc}>4$ for both doping levels $\delta=1/12$ and $1/9$. We have also measured other types of SC correlations $\Phi_{\alpha\beta}(r)$ which are provided in the SM. While they are slightly weaker than $\Phi_{yy}(r)$ due to the broken symmetry induced by the cylindrical geometry, they have very similar decaying behavior with $\Phi_{yy}(r)$. 
Although the SC correlations can be fitted by either functions, it is clear that the SC susceptibility will not diverge in the thermodynamic limit.

The spatial oscillation of the SC correlation $\Phi_{yy}(r)$ is characterized by the normalized correlation $\phi_{yy}(r)=\Phi_{yy}(r)/f(r)$, as shown in Fig.\ref{Fig:PDW}c. It is clear that $\phi_{yy}(r)$ varies periodically in real space and can be well fitted by $\phi_{yy}(r)\sim {\rm sin}(Q r+ \theta)$ for both cases. This is consistent with the PDW state with vanishing spatial average of $\phi_{yy}(r)$. The ordering wavevector of the PDW is $Q=3\pi\delta/2a_0$ as indicated by the peak position of the Fourier transformation $\phi_{yy}(q)$ of $\phi_{yy}(r)$ (Fig.\ref{Fig:PDW}d) and $\theta$ is a fitting phase factor. The corresponding wavelength is $\lambda_{sc}=2a_0/3\delta$, which is $\lambda_{sc}=8a_0$ for $\delta=1/12$ and $\lambda_{sc}=6a_0$ for $\delta=1/9$. As we will see below that the relation $\lambda_{sc}=\lambda_s=2\lambda_c$ can be clearly seen in our results as expected from that of striped PDW. Here, $\lambda_c$ and $\lambda_s$ are the wavelengths of the CDW and SDW, respectively. According to Ginzburg-Landau theory, the development of $K=2Q$ charge oscillation corresponds to the cubic term $\rho_K\Delta^*_Q \Delta_{-Q}$ in the free energy, where $\rho_K$ and $\Delta_Q$ are the charge density and PDW order parameters with corresponding momenta $K$ and $Q$, respectively. This is distinct from the CDW modulated superconductivity with term $\rho_Q\Delta^*_0 \Delta_{-Q}$ with coexisting dominant uniform $\Delta_0$ and secondary stripe pairing $\Delta_Q$. 

To identify the pairing symmetry, we have calculated the SC correlations using both real-valued and complex-valued DMRG simulations. We first rule out the $d\pm id$-wave symmetry as we find that both the wavefunction and SC correlations are real while their imaginary parts are zero. We have further analyzed the relative phase of different SC correlations, e.g., $\Phi_{xy}(r)/\Phi_{yy}(r)$ in the insets of Fig.\ref{Fig:PDW}a-b, where a clear out of phase can be observed as $\Phi_{xy}(r)/\Phi_{yy}(r)<0$. Therefore, we conclude that the pairing symmetry of the PDW is consistent with $d$-wave.

\begin{figure}[tb]
 \centering
    \includegraphics[width=1\linewidth]{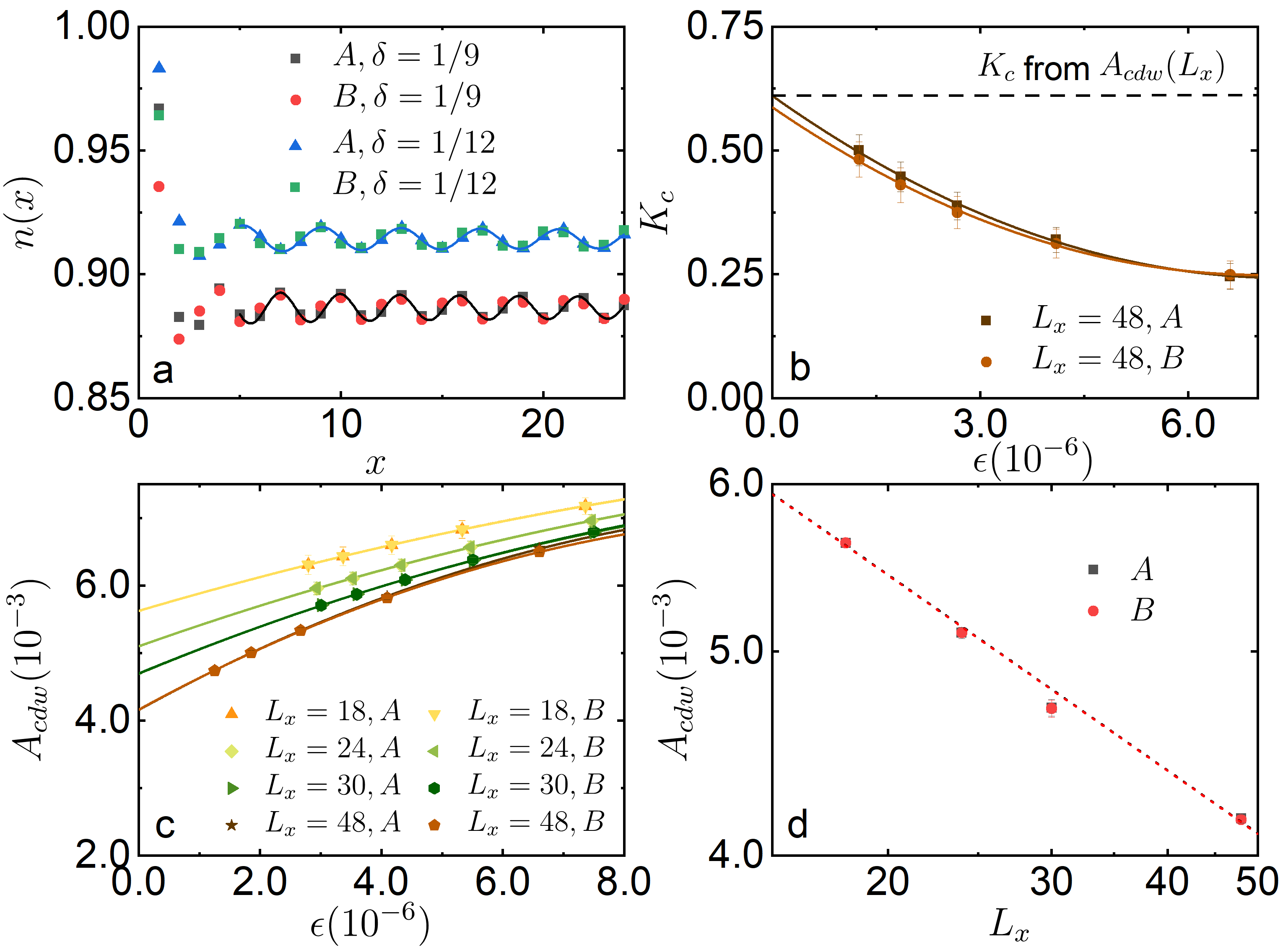}
    \caption{(Color online) (a) Charge density profiles $n(x)$ for three-leg cylinder of length $L_x=48$ at doping levels $\delta=1/9$ and $\delta=1/12$ on $A$ and $B$ sub-lattices, respectively. The solid lines denote the fitting using Eq.(\ref{Eq:Friedel}). (b) The extracted exponent $K_c$ at $\delta=1/9$ as a function of truncation error $\epsilon$. The dashed line denotes $K_c$ extracted from $A_{cdw}(L_x)$ in (d). (c) Convergence and length dependence of the CDW amplitude $A_{cdw}$ for three-leg cylinder at $\delta=1/9$ as a function of truncation error $\epsilon$. The solid lines denote fitting using second-order polynomial function. (d) Finite-size scaling of $A_{cdw}(L_x)$ as a function of $L_x$ in a double-logarithmic plot at $\delta=1/9$, where the dashed line denotes the fitting to a power law $A_{cdw}(L_x)\sim L_x^{-K_c/2}.$}\label{Fig:CDW}
\end{figure}

\begin{table}[tb]
\centering 
\begin{tabular}{C{0.15\linewidth} | C{0.15\linewidth} C{0.15\linewidth} C{0.15\linewidth} C{0.15\linewidth} C{0.15\linewidth}} 
\hline\hline 
$\delta$ & $K_c$ & $\xi_{sc}$ & $\xi_{s}$ &  $\xi_{G}$ & $c$ \\ [.5ex] 
\hline 
$1/9$   &  $0.61(3)$ & $3.1(1)$ & 1.0(1) & 1.9(1) & $\sim1$ \\
$1/12$  &  $0.62(3)$ & $3.4(3)$ & 1.2(1) & 1.8(1) & $\sim1$ \\
\hline\hline 
\end{tabular}
\caption{The table lists the Luttinger exponent $K_c$, correlation length (in unit of $a_0$) $\xi_{sc}$, $\xi_s$ and $\xi_G$, as well as central charge $c$ of the $t$-$J_K$ model at different doping levels $\delta$ in the limit $L_x=\infty$. Note that $K_c$ is obtained by fitting $A_{cdw}$ as a function of $L_x$ shown in Fig.\ref{Fig:CDW}d. }\label{Table:Exponent}
\end{table}

\textbf{Charge density wave: }%
In addition to SC correlations, we have also measured the charge density profiles to describe the charge density properties of the system. The rung charge density $n(x)=\sum_{y=1}^{L_y}n(x,y)/L_y$ is shown in Fig.\ref{Fig:CDW}, where $x$ is the rung index of the cylinder and $n(x,y)$ is the local charge density on site $i=(x,y)$. It is clear that the charge density varies periodically in real space along the $\mathbf{e}_1$ direction with the wavelength $\lambda_c=a_0/3\delta$, i.e., $\lambda_c=4a_0$ and $\lambda_c=3a_0$ for $\delta=1/12$ and $\delta=1/9$, respectively. Therefore, there are two thirds of a doped hole in each CDW unit cell. Moreover, it is clear that the relation $\lambda_{sc}=2\lambda_c$ holds for both $\delta=1/12$ and $\delta=1/9$, which is consistent the striped PDW state.
Different with SC correlations, we find power-law decay of the charge density correlation at long distances. 
The Luttinger exponent $K_c$ of the power-law decay can be extracted by fitting the charge density oscillation (Friedel oscillation) induced by the open boundaries of the cylinder\cite{White2002}%
\begin{eqnarray}
n(x)=n_0+\delta n*\cos(K x+\theta)x^{-K_c/2}. \label{Eq:Friedel}
\end{eqnarray}
Here $x$ is the distance in the $\mathbf{e}_1$ direction from the open boundary, $n_0$ the average density and $K$ is the ordering wavevector. 
$\delta n$ and $\theta$ are the model dependent constants. Examples of the fitting using Eq.(\ref{Eq:Friedel}) on the a sub-lattice are given in Fig.\ref{Fig:CDW}a for both doping $\delta=1/12$ and $\delta=1/9$, where four data points near the boundary are removed to minimize boundary effect for a more reliable fit. The extracted exponent $K_c$ is given in Table \ref{Table:Exponent}. It is clear that $K_c<1$ which demonstrates the dominance of the charge density correlations.

Alternatively, we can estimate $K_c$ from the amplitude $A_{cdw}(L_x)$ of the charge density modulation.\cite{Jiang2019hub} For a given cylinder of length $L_x$, the CDW amplitude $A_{cdw}(L_x)$ can be obtained by fitting the central-half region of the charge density profile $n(x)$.\cite{Jiang2018tJ,Jiang2019hub,Jiang2020hub} For quasi-long-range charge order, the amplitude should follow $A_{cdw}(L_x)\propto L_x^{-K_c/2}$ with similar $K_c$ with that obtained from Friedel oscillation in Eq.(\ref{Eq:Friedel}). This is indeed the case as shown in Fig.\ref{Fig:CDW}c-d, where the exponent $K_c$ is given in Fig.\ref{Fig:CDW}b and Table \ref{Table:Exponent}.

\begin{figure}[tb]
\centering
  \includegraphics[width=1\linewidth]{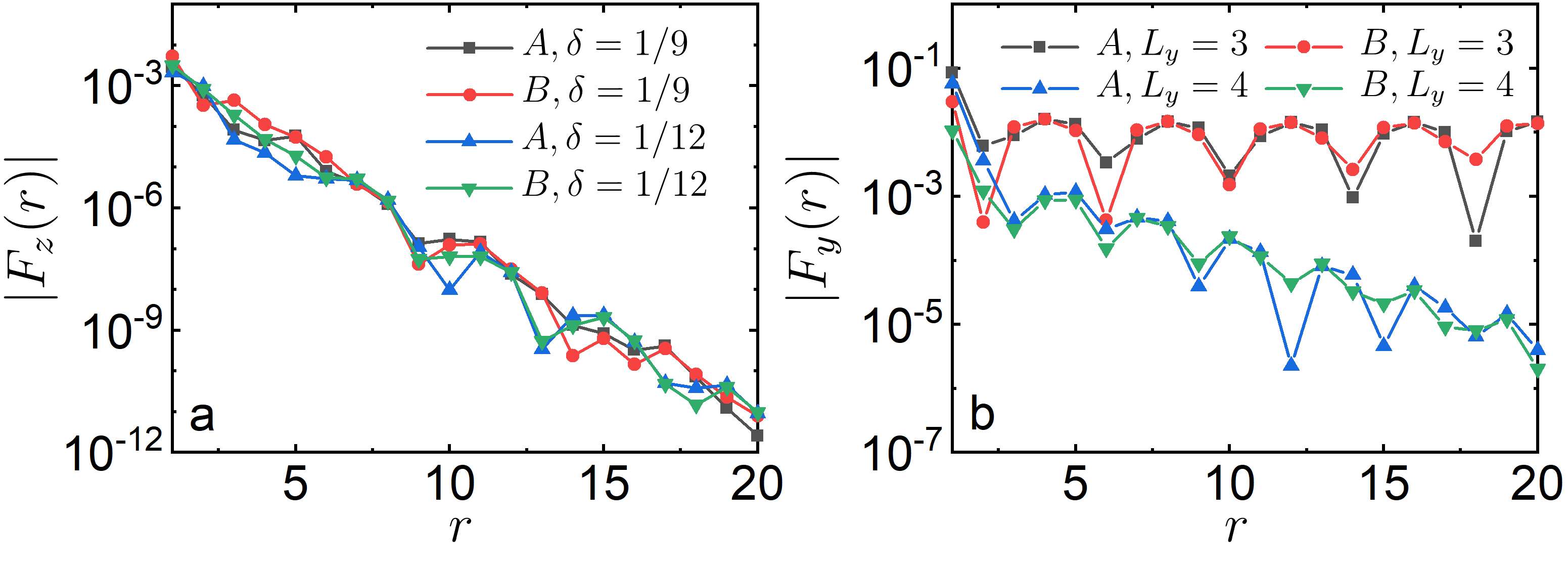}
\caption{(Color online) Spin-spin correlations (a) $F_z(r)$ for three-leg cylinder at different doping levels $\delta$ and (b) $F_y(r)$ for both three and four-leg cylinders at doping level $\delta=1/12$.}\label{Fig:Spincor}
\end{figure}

\textbf{Spin-spin and single-particle correlations: }%
To describe the magnetic properties of the system, we have further calculated the spin-spin correlation function $F_\gamma(r)=\langle S^\gamma_{i_0} S^\gamma_{i_0+r}\rangle$. Here $r$ is the distance between two sites in the $\mathbf{e}_1$ direction and $\gamma=x,y,z$. $i_0=(x_0,y)$ is the reference site with $x_0\sim \tilde{L}_x/4$. For the pure Kitaev model without doping, it is known that $F_\gamma (r)$ is nonzero only for NN sites.\cite{Kitaev2006,Baskaran2007} Upon doping, the $Z_2$ flux on each hexagonal plaquette is no longer a conserved quantity, and the spin-spin correlation functions become nonzero even at long distance. For both doping levels $\delta=1/12$ and $\delta=1/9$ on three-leg cylinders, we find that both $F_x(r)$ and $F_z(r)$ decay exponentially as $F(r)\sim e^{-r/\xi_s}$, where the spin-spin correlation length is given in Table.\ref{Table:Exponent}. On the contrary, $F_y(r)$ appears to be long-range ordered as shown in Fig.\ref{Fig:Spincor}. This is unexpected but allowed theoretically since there is no continuous spin symmetry in the system. Moreover, we find that all spin-spin correlations show clear spatial oscillation with a wavelength $\lambda_s$ that is the same as that of the SC correlation, i.e., $\lambda_s=\lambda_{sc}$, which is consistent with the striped PDW. However, our results suggest that the long-range correlation $F_y(r)$ is special to three-leg cylinder. On the contrary, $F_y(r)$ decays exponentially on the wider four-leg cylinders (Fig.\ref{Fig:Spincor}b), which is similar with both $F_x(r)$ and $F_z(r)$. However, the evidences of the striped PDW are robust for both three and four-leg cylinders with similar sign-changing SC correlations (see SM for details).

In addition to superconductivity, we have also measured the single-particle Green function $G_\sigma(r)=\langle c^{\dagger}_i c_{i+r}\rangle$ to test the possibility of Fermi liquid state.\cite{Mei2012} It is clear that the single-particle Green function decays exponentially at long distances as $G_\sigma(r)\sim e^{-r/\xi_G}$ (see SM for details). Here $\xi_G$ is the correlation length which is summarized in Table \ref{Table:Exponent}. As a result, our study suggests that the Fermi liquid state is unlikely in the lightly hole-doped KSL.

\begin{figure}[tb]
\centering
    \includegraphics[width=1\linewidth]{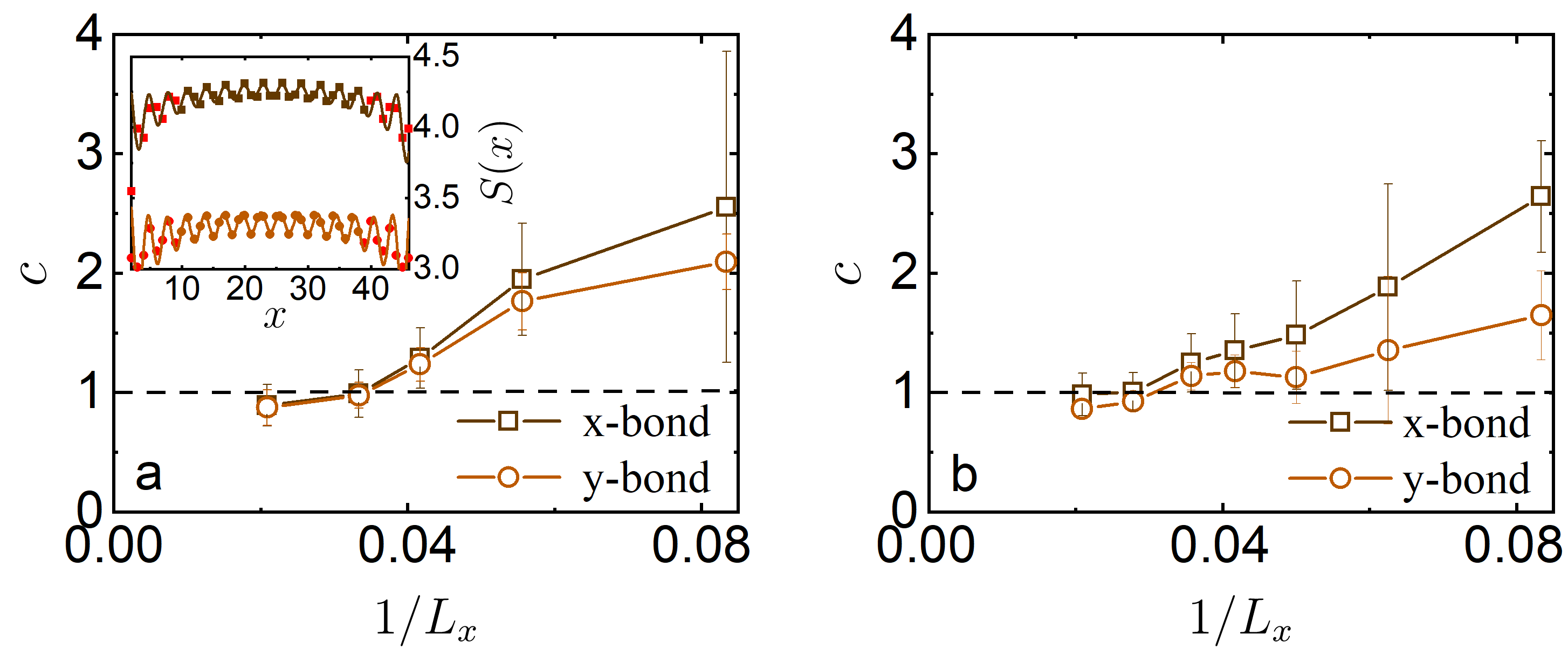}
\caption{(Color online) The extracted central charge $c$ as a function of $L_x$ using Eq.(\ref{Eq:CC}) for three-leg cylinder at doping level (a) $\delta=1/9$ and (b) $\delta=1/12$. Inset shows the entanglement entropy $S(x)$ for three-leg cylinder of length $L_x=48$ at $\delta=1/9$ with smooth boundary through $x$ and $y$-bond in the $\mathbf{e}_2$ direction, respectively. A few data points in red are removed to minimize the boundary effect. Error bars denote numerical uncertainty.} \label{Fig:Entropy}
\end{figure}

\textbf{Central charge: }%
Our results suggest that the ground state of the lightly hole-doped KSL has quasi-long-range CDW correlation with gapless charge mode. To show this, we have calculated the von Neumann entanglement entropy $S(x)=-\trace{\rho_x \ln \rho_x}$, 
where $\rho_x$ is the reduced density matrix of subsystem with length $x$. For 1D critical system described by conformal field theory, it has been established\cite{Calabrese2004,Fagotti2011} that for an open finite system of length $L_x$, %
\begin{eqnarray}
  S(x) &=& \frac{c}{6}\frac{\ln[4(L_x+1)]}{\pi} \sin[\frac{\pi(2x+1)}{2(L_x+1)}|\sin(k_F)|]\nonumber \\
   &+& \frac{a\pi \sin[k_F(2x+1)]}{4(L_x+1)\sin[\frac{\pi(2x+1)}{2(L_x+1)}|\sin(k_F)|]}+ \tilde{c}. \label{Eq:CC}
\end{eqnarray}
Here $c$ is central charge, $k_F$ denotes the Fermi momentum, $a$ and $\tilde{c}$ are model dependent parameters. Fig.\ref{Fig:Entropy}a shows an example of $S(x)$ for three-leg cylinder at doping level $\delta=1/9$. Here we have calculated $S(x)$ by dividing the system into two parts with smooth boundary through both $x$ and $y$ bonds. The extracted central charge $c$ is given in Fig.\ref{Fig:Entropy} as a function of $L_x$ at doping levels $\delta=1/9$ and $1/12$. It is clear that the central charge quickly converges to $c=1$ with the increase of $L_x$, although it deviates notably from $c=1$ on short cylinders due to finite-size effect. Therefore, our results show that there is one gapless charge mode which is consistent with the quasi-long-range CDW correlations. We have obtained similar results on four-leg cylinders which are provided in the SM.

\textbf{Summary and discussion: }%
Admittedly, the DMRG calcualtions are carried out on finite length cylinders. However, based on the results we have obtained we conjecture that the exact ground state for an infinite long three-leg cylinder has the following properties:  1) There is a single gapless charge mode and a gap (which produces exponentially falling correlations) for all spin carrying excitations. 2) Long-range SDW order with the ordered moment oriented in the $y$ direction - that is along the circumference of the cylinder - and an ordering vector $Q= 3\pi \delta/2$;  the connected spin correlations fall exponentially with a correlation length $\xi_s \sim  a_0$.  3) There are power-law CDW correlations with an exponent $K_c \sim 2/3$ and an ordering vector $K_c=2Q$. 4) There are strong even-parity d-wave-like PDW correlations with an ordering vector $Q$ which fall either exponentially with a correlation length $\xi_{sc} \sim 3a_0$ or possibly with a high power law $K_{sc} \geq 4$. 5) All other forms of SC correlations - those corresponding to odd-parity pairing or spatially uniform even-parity pairing - are much weaker in comparison with the PDW correlations.

There are many aspects of these observations that are surprising, and will need to be understood theoretically. The PDW correlations are sufficiently short-ranged that one would infer (based on any reasonable conjecture concerning their time dependence) that the corresponding PDW susceptibility would be finite. Thus, at present, we can only conclude that the present results are suggestive of a possible PDW ordered state for the 2D (infinite leg) version of this model. However, it is notable that the favored forms of order are remarkably reminiscent of those conjectured to be present in the cuprate high temperature superconductor, LBCO, where direct evidence exists of stripe SDW order with wave vector $Q \approx 4\pi /\delta a$ and CDW order with ordering vector $K = Q/2$, and indirect evidence has been adduced for PDW order with ordering vector $Q$.

In this paper, we primarily focus on the lightly doped Kitaev model, it will be interesting to study the higher doping case as well as the extend Kitaev model with further neighbor hopping which is shown to be essential to enhance the superconductivity on the square lattice.\cite{Jiang2019hub,Jiang2020hub} As other terms such as the Heisenberg interaction and spin-orbit couplings are also present in real materials such as ${\rm Na_2IrO_3}$ and $\alpha$-${\rm RuCl_3}$,\cite{Jackeli2009,Jiang2011,Singh2012,Chun2015,Kim2016,Banerjee2017,Trebst2017,Takagi2019,Hickey2019,Jiang2019kh,Yokoi2020} it will be interesting to study these systems as well.

{\it Acknowledgments:} We would like to thank Senthil
Todadri for insightful discussions, and especially Steve Kivelson for insightful discussion, invaluable suggestions and generous help to understand the results and improve the manuscript. This work was supported by the Department of Energy, Office of Science, Basic Energy Sciences, Materials Sciences and Engineering Division, under Contract DE-AC02-76SF00515. Parts of the computing for this project was performed on the Sherlock cluster.

\bibliography{Refs}

\clearpage

\appendix

\setcounter{equation}{0}
\setcounter{figure}{0}
\setcounter{table}{0}
\setcounter{page}{1}
\makeatletter
\renewcommand{\theequation}{A\arabic{equation}}
\renewcommand{\thefigure}{A\arabic{figure}}

\section{Supplemental Material}

\subsection{Results for four-leg cylinders}
\begin{figure}[tb]
\centering
    \includegraphics[width=1\linewidth]{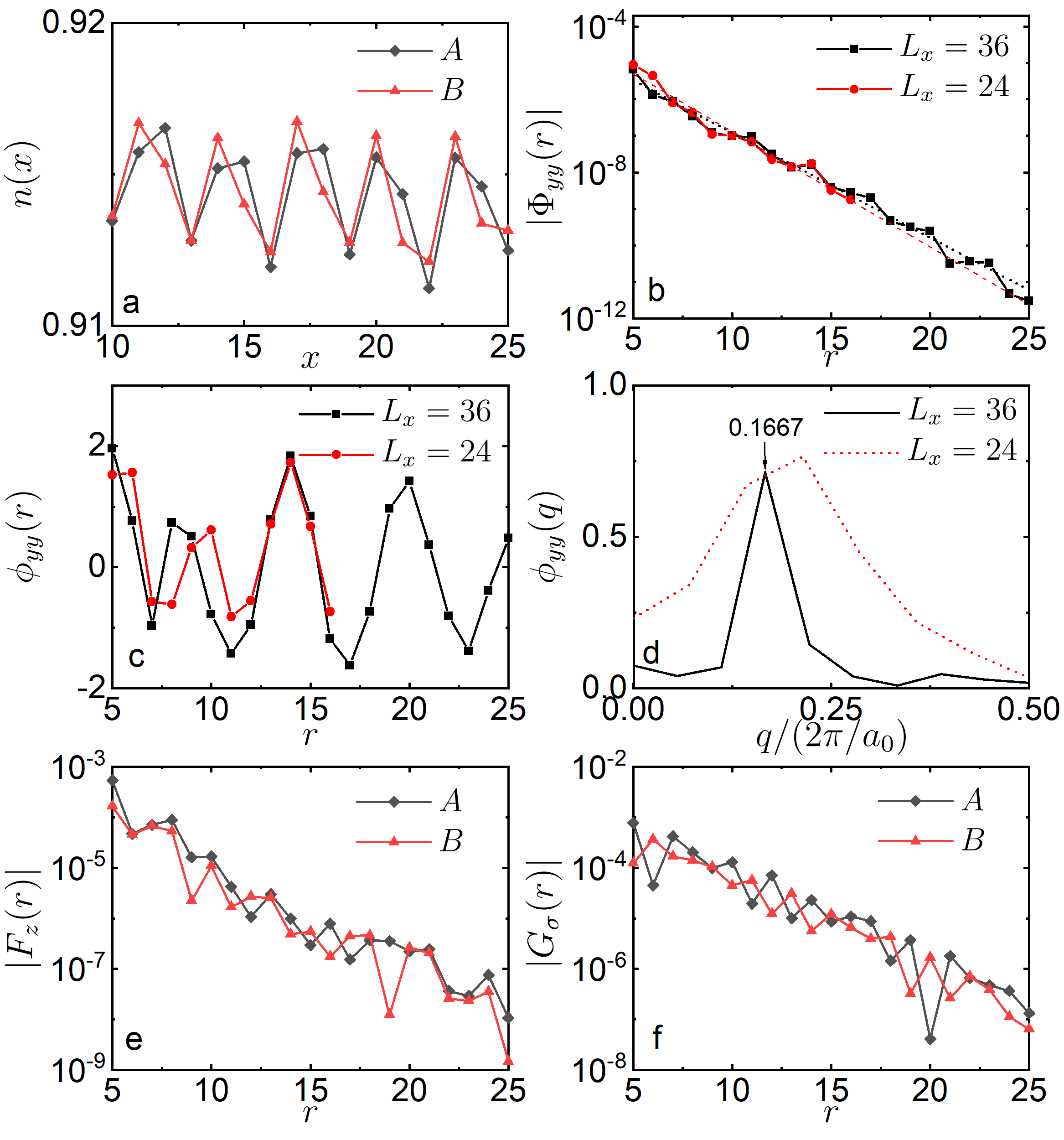}
\caption{(Color online) Physical quantities on four-leg cylinder of length $L_x=24\sim 36$ at doping $\delta=1/12$. (a) Charge density profile $n(x)$ on A and B sub-lattices, respectively. (b) The SC pair-pair correlation $|\Phi_{yy}(r)|$. The dashed lines label the fitting using the exponential function $f(r)\sim e^{-r/\xi_{sc}}$. (c) The normalized function $\phi_{yy}(r)=\Phi_{yy}/f(r)$. (d) Fourier transformed $\phi_{yy}(q)$. (e) The spin-spin correlation $F_z(r)$ on A and B sub-lattices, respectively. (f) The single-particle Green's function $G_{\sigma}(r)$.} \label{Apdx:SC_4leg}
\end{figure}

We provide more results for four-leg cylinder, where we have observed similar features of the striped PDW discussed in the main text. Similar with three-leg cylinders, the density profiles on four-leg cylinders also exhibit similar spatial oscillation (Fig.\ref{Apdx:SC_4leg}a), with the period $\lambda_c=a_0/4\delta$ corresponding to an ordering vector $K=4\pi\delta/a_0$, where $a_0$ is the length of unit cell.

The SC pair-pair correlation function $\Phi_{yy}(r)$ on four-leg cylinder of length $L_x=24$ and $L_x=36$ at $\delta=1/12$ doping are shown in Fig.\ref{Apdx:SC_4leg}b. Similar with three-leg cylinders, the SC pair-pair correlation $\Phi_{yy}(r)$ decays exponentially whose envelope can be well fitted by the function $f(r)\sim e^{-r/\xi_{sc}}$ with a correlation length $\xi_{sc}=1.37(8)$ and $\xi_{sc}=1.51(3)$ for $L_x=24$ and $L_x=36$, respectively. The spatial oscillation of $\Phi_{yy}(r)$ is characterized by the normalized pair-pair correlation $\phi_{yy}=\Phi_{yy}(r)/f(x)$ shown in Fig.\ref{Apdx:SC_4leg}c. The ordering wavevector of $\Phi_{yy}(r)$ is $Q=2\pi\delta/a_0$ as indicated by the peak position of the Fourier transformed $\phi_{yy}(q)$ in Fig.\ref{Apdx:SC_4leg}d. The relationship $Q=K/2$ between the SC ordering vector $Q$ and CDW ordering vector $K$ also holds true for four-leg cylinders, as expected for the striped PDW state. 

On four-leg cylinders, all the spin-spin correlations $F_{\gamma}(r)$, where $\gamma=x,y,z$, decay exponentially at long distances, e.g., $F_z(r)$ as shown in Fig.\ref{Apdx:SC_4leg}e. The envelope of the strongest component $F_y(r)$ (see Fig.\ref{Fig:Spincor}b) can be well fitted by the exponential function $f(r)\sim e^{-r/\xi_{s}}$ with a correlation length $\xi_{s}=2.76(8)$ from length $L_x=36$ cylinder. The single-particle Green's function also decays exponentially at long distances as $G_{\sigma}(r)\sim e^{-r/\xi_{G}}$ (Fig.\ref{Apdx:SC_4leg}f) with the correlation length $\xi_{G}=2.48(5)$.

\begin{figure}[tb]
\centering
    \includegraphics[width=1\linewidth]{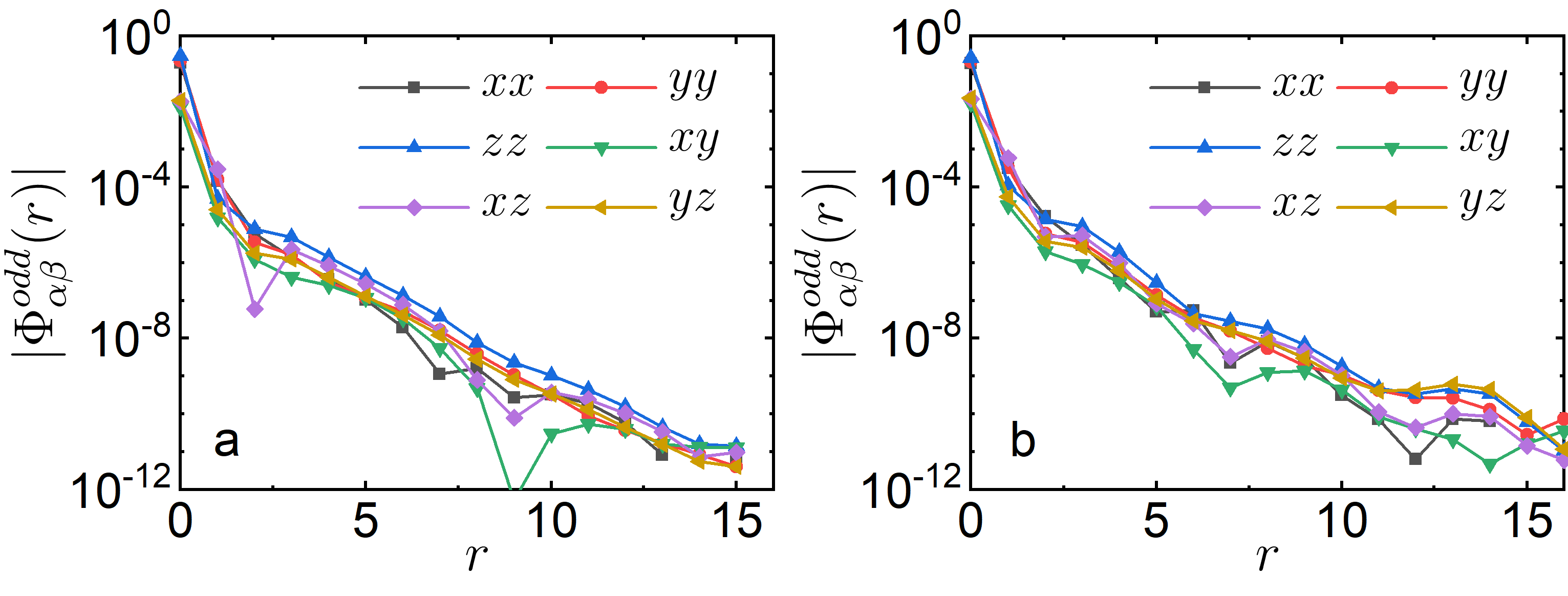}
\caption{(Color online) Various types of odd-parity SC correlations on three-leg cylinders of lengths (a) $L_x=30$ at doping $\delta=1/9$ and (b) $L_x=36$ at doping $\delta=1/12$.}\label{Apdx:Oddpair}
\end{figure}

\subsection{Odd-parity SC pair-pair correlation function}
We measure the odd parity pair-pair correlation function defined as
\begin{eqnarray}
\Phi^{odd}_{\alpha\beta}(r)=\langle\Delta^{\dagger}_{odd,\alpha}(x_0,y)\Delta_{odd,\beta}(x_0+r,y)\rangle,
\label{Eq:TC}
\end{eqnarray}
where $\Delta^{\dagger}_{odd,\alpha}(x,y)=\frac{1}{\sqrt{2}}[c^{\dagger}_{(x,y),\uparrow}c^{\dagger}_{(x,y)+\alpha,\downarrow}+c^{\dagger}_{(x,y),\downarrow}c^{\dagger}_{(x,y)+\alpha,\uparrow}]$. $\alpha=x,y,z$ labels the bond orientations as defined in the main text. Compared with the even-parity SC correlation $\Phi(r)$, $\Phi^{odd}_{\alpha\beta}(r)$ shown in Fig.\ref{Apdx:Oddpair} are much weaker which decay faster with shorter correlation length, e.g., $\xi=0.86(1)$ at $\delta=1/12$ doping and $\xi=1.09(8)$ at $\delta=1/9$ doping.

\begin{figure}[tb]
\centering
    \includegraphics[width=1\linewidth]{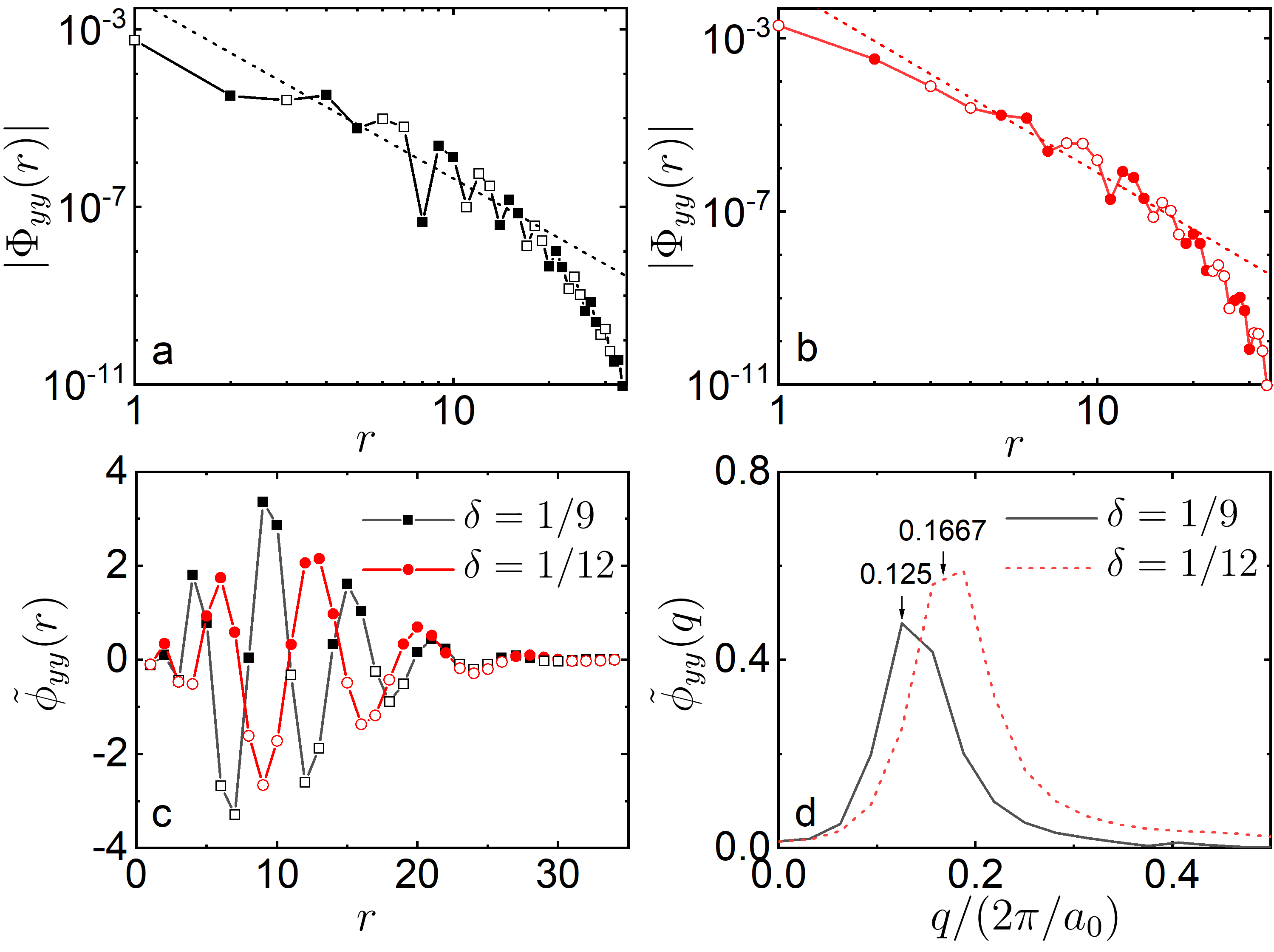}
\caption{(Color online) Power-law fitting of the SC pair-pair correlation on three-leg cylinders of length $L_x=48$ at doping (a) $\delta=1/9$ and (b) $\delta=1/12$. The filled (open) symbols denote the positive (negative) values. The dashed lines label the fitting function $\tilde{f}(x)\sim r^{-K_{sc}}$. (c) The normalized function $\tilde{\phi}_{yy}(r)=\Phi_{yy}/\tilde{f}(r)$ at doping $\delta=1/9$ and $1/12$. (d) Fourier transformed $\tilde{\phi}_{yy}(q)$ at doping $\delta=1/9$ and $1/12$, with the peak located at momentum $Q= 3\pi\delta/2a_0$.}\label{Apdx:pdwpl}
\end{figure}

\subsection{Power-law fitting of SC pair-pair correlation}
Here we double-check the alternative power-law fitting for the SC pair-pair correlation. Fig.\ref{Apdx:pdwpl} shows the same SC pair-pair correlations $\Phi_{yy}(r)$, i.e., Fig.\ref{Fig:PDW} in the main text, in double-logarithm scale. At long distances, it is quite clear that $\Phi_{yy}(r)$ deviates from the power-law fitting labelled by the linear dashed lines $\tilde{f}(r)\sim r^{-K_{sc}}$ with the exponent $K_{sc}=4.0(4)$ and $4.3(2)$ for $\delta=1/9$ and $\delta=1/12$ dopings, respectively. This deviation is also reflected in the vanishing normalized correlation $\tilde{\phi}_{yy}(r)=\Phi_{yy}(r)/\tilde{f}(r)$ at long distance and the rounded peak of $\tilde{\phi}_{yy}(q)$ located at the momentum $q=3\pi\delta/2a0$ shown in Fig.\ref{Apdx:pdwpl}c and d.

\subsection{All types of even parity pair-pair correlations}
\begin{figure}[htbp!]
\centering
    \includegraphics[width=1\linewidth]{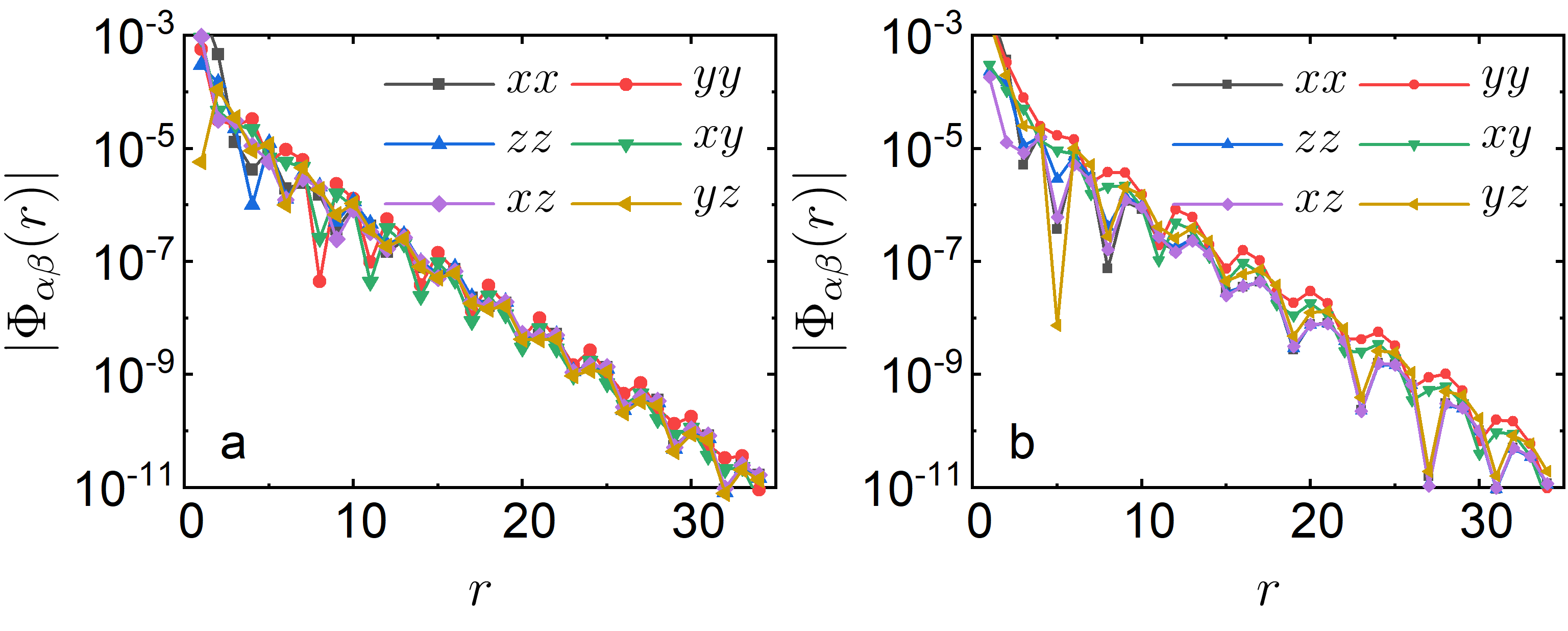}
\caption{(Color online) Various types of even-parity SC pair-pair correlations on three-leg cylinders of length $L_x=48$ at doping (a) $\delta=1/9$ and (b) $\delta=1/12$. }
\label{Apdx:SC_3leg}
\end{figure}

In Fig.\ref{Apdx:SC_3leg}, we list all types of even-parity SC pair-pair correlations $\Phi_{\alpha\beta}(r)$ on three-leg cylinders of length $L_x=48$ at doping $\delta=1/9$ and $1/12$. It is clear that they all decay exponentially, similar with $\Phi_{yy}(r)$ discussed in main text, although with slightly different amplitude at long distances. This could be attributed to the fact that the cylinder geometry breaks the $C_3$ rotational symmetry of the honeycomb lattice.

\subsection{Single-particle Green function}%
\begin{figure}[tb]
 \centering
   \includegraphics[width=1\linewidth]{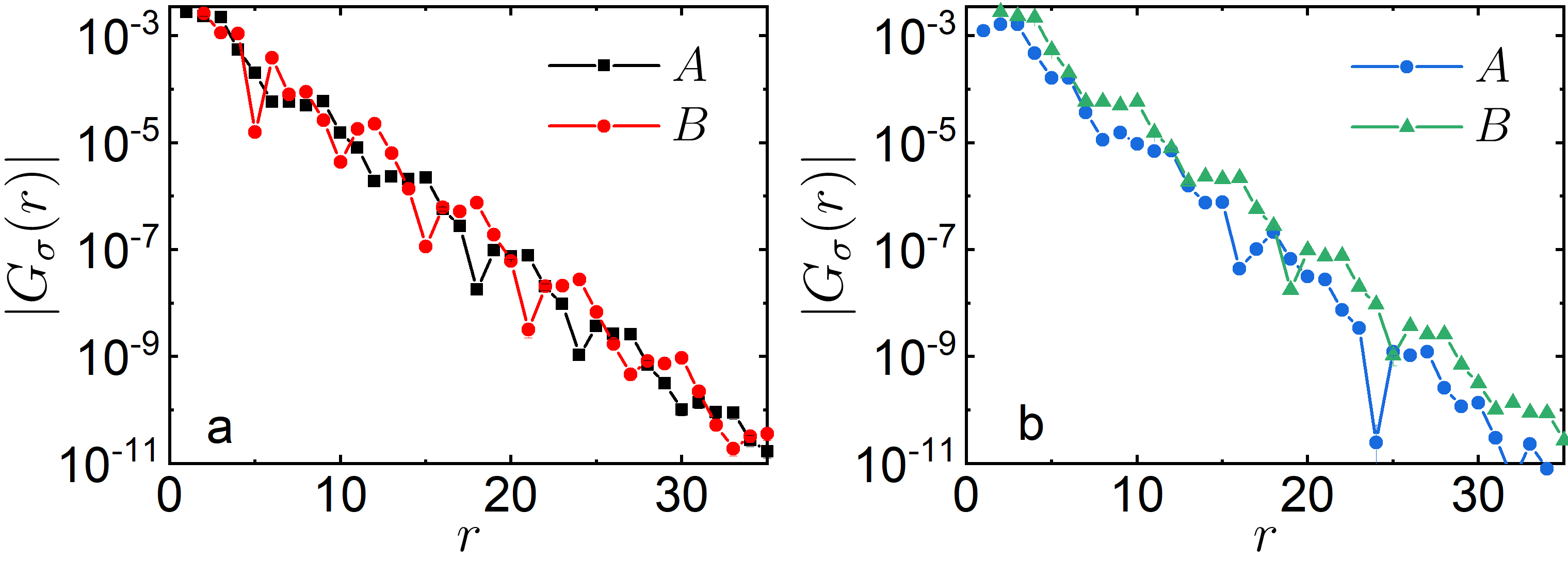}
\caption{(Color online) Single-particle Green function $G_{\sigma}(r)$ on three-leg cylinders of length $L_x=48$ at doping (a) $\delta=1/9$ (b) $\delta=1/12$.}
\label{Apdx:siglp}
\end{figure}

To check the possibility of other metallic state such as the Fermi liquid state, We have also measured the single-particle Green function $G_\sigma (r)$, as shown in Fig.\ref{Apdx:siglp}.
It is clear that $G_{\sigma}(r)\sim e^{-r/\xi_G}$ decays exponentially, which is inconsistent with the Fermi liquid state. For instance, the correlation lengths at doping $\delta=1/9$ and $1/12$ are $\xi_G=1.7(1)$ and $1.71(3)$ on three-leg cylinders of length $L_x=48$. The correlation length $\xi_G$ in the long-cylinder limit after finite-size scaling using $L_x= 18\sim 48$ is slightly longer, which is given in Table I of the main text.

\end{document}